\documentclass[preprint,aps,nofootinbib]{revtex4}
\usepackage{float}
\usepackage{graphicx}
\input{epsf.sty}
\usepackage{epsfig}

\def\lsim{\mathrel{\raise.3ex\hbox{$<$\kern-.75em\lower1ex\hbox{$\sim$}}}}
\def\gsim{\mathrel{\raise.3ex\hbox{$>$\kern-.75em\lower1ex\hbox{$\sim$}}}}

\def\singleandabitspaced{\baselineskip=\normalbaselineskip\multiply
    \baselineskip by 150\divide\baselineskip by 100}

\begin{document}

%\singlespaced

%\hfill$\vcenter{\hbox{\bf hep-ph/0701259}}$
\vskip 0.4cm

\title{Aspects of the Screening Length and Drag Force in Two Alternative Gravity Duals of the Quark-gluon Plasma \\
} 
\author{Piyabut Burikham \footnote{piyabut@iastate.edu}, Jun Li \footnote{junli@iastate.edu}}
\vspace*{0.5cm}
\affiliation{Department of Physics and Astronomy, Iowa State University, \\
        Ames, Iowa 50011, USA} 
 
\date{\today}

\vspace*{2.0cm}

\begin{abstract}
We compute the screening length of mesons with different angular momentum $J$ from two gravity 
dual theories.  Both the asymptotically $AdS_{5}$ and Sakai-Sugimoto metrics are considered in the calculations.  Using the dual description of the quark as a classical string ending on the probe brane, we obtain the interacting potential between the heavy quark and antiquark after rotating the background metric.  The result shows that the screening length of mesons with different $J$ is well fit to $a_{J}/T$.  The constant $a_{J}$ is determined for $J=0,1,2$ by taking advantage of numerical techniques.  Finally, we calculate the drag force and relaxation times from the Sakai-Sugimoto metric and compare with the ones obtained in the $AdS_{5}$.  The application of our result to charmonium and bottomonium at RHIC and LHC is briefly discussed.

\end{abstract}

\maketitle

%\end{titlepage}

\newpage

\setcounter{page}{2}
\renewcommand{\thefootnote}{\arabic{footnote}}
\setcounter{footnote}{0}
\singleandabitspaced

\section{Introduction}

The Relativistic Heavy Ion Collider (RHIC) at Brookhaven National Laboratory has announced that there is a new phase of matter, the quark-gluon plasma, produced in the Au-Au collision.  Experimental results from RHIC indicate that this new phase of nuclear matter has a small shear viscosity per entropy density ratio $\eta/s$ near $T_{c}$ (phase transition temperature from hadronic matter to quark-gluon plasma).  This implies the strongly coupled quark-gluon plasma was formed near $T_{c}$~\cite{gm}.  One can investigate its thermal properties based on the probes which provide us with the information about the presence of this nuclear matter.  One of the tools is the quarkonium.  Quarkonium is bound state of the heavy quark and antiquark pair.   To estimate the energy and radius of these bound states, one can solve the Schr\"{o}dinger equation with the non-relativistic empirical potential~\cite{ee}
\begin{eqnarray}
V(r,T=0)& = & \sigma r-\frac{\alpha_{\mbox{eff}}}{r}\label{eq:clpot1}
\end{eqnarray}
where $\sigma$ is the QCD-string tension and $\alpha_{\mbox{eff}}$ is the effective coupling.
In the hot quark-gluon plasma, the deconfined  charges would modify the interaction between the heavy quark pair such that we expect the colour-screened potential~\cite{kms}
\begin{eqnarray}
V(r,T)& = & (\sigma r_{D}(T))\left( 1-e^{-r/r_{D}(T)}\right) - \frac{\alpha_{\mbox{eff}}}{r}e^{-r/r_{D}(T)}\label{eq:clpot2} 
\label{eq:clpot3}
\end{eqnarray}      
where $r_{D}(T)$ is the screening radius.  When the temperature increases, the screening radius $r_{D}(T)$ decreases.  Roughly speaking, once the screening radius $r_{D}(T)$ is smaller than the radius of bound state of the heavy quark pair, the heavy quark and antiquark would not bind together and the quarkonium dissociates.  At RHIC, the relatively light charmonia such as $J/\psi$ could be produced in a large number in the initial stage of the collision process.  If there is a quark-gluon plasma formed in the region of produced charmonium, it will affect the production such that the yield of $J/\psi$ will be suppressed as compared with the yield without the quark-gluon plasma.  Thus the strong suppression of $J/\psi$ production can be used as the indication of the formation of the quark-gluon plasma~\cite{ms}\footnotemark[1]\footnotetext[1]{whether the direct $J/\psi$ production is suppressed at RHIC or not is still unclear provided that current lattice results show the surprising resistance to the melting of the 1S states such as $J/\psi,\eta_{c}$ at least until $T>1.5T_{c}$.  The $\psi^{\prime}$(2S),$\chi_{c}$(1P), however, do melt right above $T_{c}$ even in the lattice determination~\cite{lat,sat2}.  Due to the suppressed sequential decay $\psi^{\prime},\chi_{c}\to J/\psi$, the indirect $40\%$ suppression of $J/\psi$ production could still exist in this scenario. }.  The screening radius is a good indicator of the presence of the quark-gluon plasma, as well as a good probe to estimate the temperature profile of the produced quark-gluon nuclear phases.    

The screening radius $r_{D}(T)$ estimated by the lattice calculations gives $r_{D}(T)\simeq 0.21\sim0.24$ fm at $T=250$ MeV for $T_{c}=175\sim190$ MeV~(from Fig.~11 of Ref.~\cite{kz}).  Remarkably, the screening length calculated by the gravity dual theory in the original AdS/CFT correspondence~\cite{mal} for zero-spin meson gives the value $0.19$ fm at $T=250$ MeV~\cite{gub1}.  Even this is close to the lattice result, it is desirable to take into account the nonzero angular momentum of the quark antiquark pair in the calculation.  One way to incorporate the angular momentum of the mesonic states is to rotate the background metric of the corresponding gravity dual theory~\cite{rot}.  We will adopt this approach in this article.    

%In the gravity dual picture, the screening length calculation can be performed by solving for the distance at which the regulated potential~(the difference between the potential of the connected string configuration and the parallel strings configuration) generated between the quark antiquark pair vanishes.  For a given metric, the calculation can be done by solving for the classical solution and calculate the corresponding Nambu-Goto action.       

The original AdS/CFT correspondence proposes that the $\mathcal{N}=4$ super Yang-Mills (SYM) theory at large $N_{c}$ in four dimensions is dual to the low energy super gravity approximation to type II$B$ string compactified on $AdS_{5} \times S^{5}$~\cite{mal}.  Because of the conformal invariance and supersymmetry, $\mathcal{N}=4$ SYM is not exactly equivalent to QCD.  In QCD, the coupling constant is the function of momentum transfer with a non-zero $\beta$ function and quark is in the fundamental representation.  In $\mathcal{N}=4$ SYM, the coupling constant is constant ($\beta$ function is equal to zero) and all the particles are in the adjoint representation.  Even though SYM is fundamentally different from the real QCD, they both have certain similar properties of the strongly coupled non-Abelian gauge plasma.  The calculation of ratio of the shear viscosity divided by entropy density is equal to $1/4\pi$ in SYM.  This result is taken to be the lower bound of the strongly coupled quark-gluon plasma~\cite{dts}.  This gravity dual, however, does not contain chiral quarks and the corresponding chiral symmetry restoration transition.  An {\it ad hoc} way of introducing heavy quark into the theory is to use probe ``flavor'' branes sitting at certain position in the radial direction.

As an alternative gravity dual to the strongly interacting QCD plasma, we also consider Sakai-Sugimoto model~\cite{ss} in this article.  The dual gauge theory is the maximally supersymmetric gauge theory in 5 dimensions~(with one dimension compactified).  It contains chiral quarks and a natural description of the deconfinement and chiral symmetry breaking transitions.  The open-string excitations connecting the (anti-)D8-branes to the D4-branes are identified as heavy quarks.                

Using this model, we recalculate the screening lengths incorporating angular momentum of the classical string, and compare with the ones obtained from the $AdS_{5}\times S_{5}$ gravitational background.  A few comments on drag force and energy loss in this gravity dual and in typical $AdS_{5}$ are made.  We note the different temperature dependences due to the dimensional 't~Hooft coupling in 5 dimensional gauge theory.

The article is organized as the following.  In Section II, we describe the general setup for both gravity duals, and write down the equation of motion for the classical rotating connected-string configuration.  In Section III, we numerically estimate the screening lengths in both gravity duals and discuss the characteristics of the meson ``melting'' in the quark-gluon plasma.  Finally, calculations of the drag force, energy loss, and relaxation times have been performed in Section IV.  Section V is the Conclusions and Discussions.   

%Each dual gravitational picture gives description of bound states with different $J$ in the quark-gluon plasma.  Due to the larger radius of excited states of quarkonium,  they are expected to melt earlier than the ground state of quarkonium.  Eqn.~(\ref{eq:Lw3}) and Eqn.~(\ref{eq:Lw4}) can be used to determine the temperature at which the excited bound states begin to melt.

%At the Large Hadron Collider (LHC), the higher energies will provide the opportunity to investigate the properties of $b\bar{b}$ bound states such as $\Upsilon$ states which have the smaller radii.  In section, the melting temperaure of $\Upsilon$ states are estimated provided the radius of $\Upsilon$ is given.  In section, we calculate the drag force using Sakai-Sugimoto model and compare it with the drag force from $AdS_{5}\times S_{5}$ gravitational background.  The difference between these two model are discussed.     

\section{General Setup}
To calculate the potential between the static quark and antiquark pair at the finite temperature, one can use a Schwarzschild-anti-de Sitter Type II$B$ supergravity compactification~\cite{RTY}.  However one may consider the more general situation where the quark and antiquark are rotating in the deconfined quark-gluon plasma.  The bound state of quark and antiquark with rotation corresponds to the meson with angular momentum.  In order to calculate the potential between the rotating quark and antiquark pair, we will consider the following 5 dimensional metrics given by
\begin{eqnarray}
ds^{2}=\left(\left(\frac{u}{R_{n}}\right)^{\frac{n}{2}}\left(-f_{n}\left(u\right)dt^{2}+d\rho^{2}+\rho^{2}d\varphi^{2}+dz^{2} \right)+\left(\frac{R_{n}}{u}\right)^{\frac{n}{2}}\frac{du^{2}}{f_{n}\left(u\right)}\right)
\end{eqnarray}
where
\begin{eqnarray}
f_{n}\left(u\right)=1-\frac{u_{h}^{n}}{u^{n}}, 
\end{eqnarray}
$u$ is the radial direction, $z$ is the direction perpendicular to the plane of rotation and position of the horizon $u_{h}=\frac{16}{9}\pi^{2}R_{3}^{3}T^{2}, \pi R_{4}^{2}T$ for $n=3,4$ respectively.

The $n=3$ metric, known as Sakai-Sugimoto metric~\cite{ss}~(modulo details of the compact 5 dimensional manifold which we ignore here) in the high temperature phase~\cite{psz}, is the near-horizon limit of the metric induced by a configuration of $N_{c}$ D4-branes intersecting with $N_{f}$ D8-branes and $N_{f}$ anti-D8-branes.  The (anti-)D8-branes are located at $x_{4}=0,(L)$ of the compactified $x_{4}$ around which the D4-branes wrap.  The string theory in this background is dual to the maximally supersymmetric $SU(N_{c})$ Yang-Mills theory in $1+4$ dimensions with one dimension compactified on a circle with radius $R_{3}$.

The $n=4$ metric is the near-horizon limit of the metric induced by a configuration of $N$ D3-branes in the 1+9 dimensional background.  In this limit, the non-compact 5 dimensional subspace is approximately $AdS_{5}$.  The string theory in this background is dual to the supersymmetric $\mathcal{N}=4$ Yang-Mills in 1+3 dimensions when the other compact 5 dimensional manifold is $S_{5}$.  
The horizon in the gravity picture induces the scale and Hawking temperature into the theory.  The dual of these quantities in the gauge picture are $\Lambda_{QCD}$ and temperature of the quark-gluon plasma respectively.      

The Nambu-Goto action is given by
\begin{eqnarray}
S=\frac{1}{2\pi}\int d\tau d\sigma\sqrt{det(G_{MN}\partial_{\alpha}X^{M}\partial_{\beta}X^{N})}.
\end{eqnarray}

In order to calculate the potential and solve the equation of motion,  we assign the following {\it ansatz} for the rotating connected-string configuration,
\begin{eqnarray}
t=\tau,\,\,\,\rho=\sigma,\,\,\, u=u\left(\rho\right),\,\,\,\varphi=\omega t.
\label{eq:g1}
\end{eqnarray}

The Nambu-Goto action becomes
\begin{eqnarray}
S=\frac{1}{2\pi}\int dtd\rho\sqrt{\left(\frac{u^{\prime2}}{f_{n}\left(u\right)}+\left(\frac{u}{R_{n}}\right)^{n}\right)\left(f_{n}-\omega^{2}\rho^{2}\right)}.
\end{eqnarray}
where $u^{\prime}$ is the derivative on $\rho$.

Now we can get the potential between quark and antiquark which is  
\begin{eqnarray}
V=\frac{1}{\pi}\int d\rho\sqrt{\left(\frac{u^{\prime2}}{f_{n}\left(u\right)}+\left(\frac{u}{R_{n}}\right)^{n}\right)\left(f_{n}\left(u\right)-\omega^{2}\rho^{2}\right)}.
\label{eq:V}
\end{eqnarray}
The conserved angular momentum can be written as
\begin{eqnarray}
J=\int d\rho\frac{\omega\rho^{2}\left(\frac{u^{\prime2}}{f_{n}\left(u\right)}+\left(\frac{u}{R_{n}}\right)^{n}\right)}{\sqrt{\left(\frac{u^{\prime2}}{f_{n}\left(u\right)}+\left(\frac{u}{R_{n}}\right)^{n}\right)\left(f_{n}\left(u\right)-\omega^{2}\rho^{2}\right)}}.
\label{eq:J}
\end{eqnarray}

The equation of motion for the connected-string configuration is
\begin{equation}
\frac{d}{d\rho}\left(\frac{1}{\sqrt{\left(\frac{u^{\prime2}}{f_{n}}+\left(\frac{u}{R_{n}}\right)^{n}\right)\left(f_{n}-\omega^{2}\rho^{2}\right)}}\frac{u^{\prime}}{f_{n}}\left(f_{n}-\omega^{2}\rho^{2}\right)\right)  \\\\\
-\frac{f_{n}^{\prime}\left(\frac{u^{\prime2}}{f_{n}^{2}}\omega^{2}\rho^{2}+\left(\frac{u}{R_{n}}\right)^{n}\right)+\frac{nu^{n-1}}{R_{n}^{n}}\left(f_{n}-\omega^{2}\rho^{2}\right)}{2\sqrt{\left(\frac{u^{\prime2}}{f_{n}}+\left(\frac{u}{R_{n}}\right)^{n}\right)\left(f_{n}-\omega^{2}\rho^{2}\right)}}=0.
\label{eom}
\end{equation}
The string ending on the probe brane is vertical to brane and the position is fixed to the position of brane.  In order to solve the equation of motion numerically, we impose the boundary condition
\begin{eqnarray}
u(\rho_{0})=u_{max},\,\,\,u^{\prime}(\rho_{0})\to\infty,
\label{eq:bc}
\end{eqnarray}
where $u_{max}$ is the position of the probe brane, $\rho_{0}$ is the position of endpoints of the string.  Using the boundary conditions, we can solve this differential equation numerically for each value of $\omega$ to get the string configuration solution $u=u(\rho)$.  Then we use this solution to calculate the potential $V(\omega)$ and the angular momentum $J(\omega)$.  Notice that both the potential and angular momentum depend on the angular frequency $\omega$.  We can see that the potential between the quark and antiquark will depend on the angular momentum $J$ after we eliminate $\omega$ from Eqn.~(\ref{eq:V}) and Eqn.~(\ref{eq:J}).

\section{Characteristics of the Screening Length and Holographic Melting }
\subsection{$\omega=0$ case}

First, we will consider the simplest case when there is no rotation.  The dual Yang-Mills state in this case will be the zero angular-momentum mesonic state.  
For $\omega=0$ the action
\begin{eqnarray}
S=\frac{T}{2\pi}\int dx\sqrt{u^{\prime2}+f_{n}\left(\frac{u}{R_{n}}\right)^{n}}.
\end{eqnarray}
With the simplified action, we have one more symmetry which could simplify our analysis by the use of the corresponding conserved quantity.  Since the Lagrangian does not depend on $x$ explicitly, we have a constant of the motion
\begin{eqnarray}
A \equiv L-u^{\prime}\frac{\partial L}{\partial u^{\prime}} = \frac{f_{n}\left(\frac{u}{R_{n}}\right)^{n}}{\sqrt{u^{\prime2}+f_{n}\left(\frac{u}{R_{n}}\right)^{n}}}.
\end{eqnarray}

Then we can find the shape of the string which minimizes the action in terms of the relationship between $dx$ and $du$ as
\begin{eqnarray}
dx=\frac{du}{\sqrt{f_{n}\left(\frac{u}{R_{n}}\right)^{n}\left(\frac{f_{n}\left(\frac{u}{R_{n}}\right)^{n}}{A^{2}}-1\right)}},
\end{eqnarray}
leading to the distance between quark and antiquark
\begin{eqnarray}
\frac{L}{2}=\int_{u_{b}}^{u_{max}}\frac{R_{n}^{\frac{n}{2}}u_{0}^{\frac{n}{2}}}{\sqrt{\left(u^{n}-u_{h}^{n}\right)\left(u^{n}-u_{b}^{n}\right)}}\, du
\label{eq:L0}
\end{eqnarray}
where $u_{b}=\left(u_{h}^{n}+u_{0}^{n}\right)^{\frac{1}{n}}, u_{0}^{n}=A^{2}R_{n}^{n}$, and $u_{max}$ is position of the probe brane.
The potential, Eqn.~(\ref{eq:V}), can then be expressed as
\begin{eqnarray}
V=\frac{1}{\pi}\int_{u_{b}}^{u_{max}}\frac{\sqrt{u^{n}-u_{h}^{n}}}{\sqrt{u^{n}-u_{b}^{n}}}\, du.
\label{eq:V0}
\end{eqnarray}
In order to consider stability of the connected-string configuration dual to the mesonic state, the comparison of the energy or potential with other configuration where the strings are splitted, i.e. parallel strings configuration, has to be made.  At non-zero temperature, there will be certain distance between quark and antiquark such that the potential of the connected-string starts to be larger than the potential of the parallel strings configuration.  Therefore the distance signifies the dissociation of the quark antiquark system.  It could be identified with the screening length of the meson in the gauge plasma.

For the parallel strings configuration, we choose the following worldsheet gauge,
\begin{eqnarray}
t=\tau,\,\,\,u=\sigma,\,\,\, \rho=\rho_{0},
\label{eq:g2}
\end{eqnarray}
where $\rho_{0}$ is the position of endpoints of the string.

The corresponding regulating potential then becomes
\begin{eqnarray}
V^{\prime}=\frac{1}{\pi}\int_{u_{h}}^{u_{max}}du.
\end{eqnarray}

\begin{figure}[tb]
\centering
\includegraphics[width=3.2in]{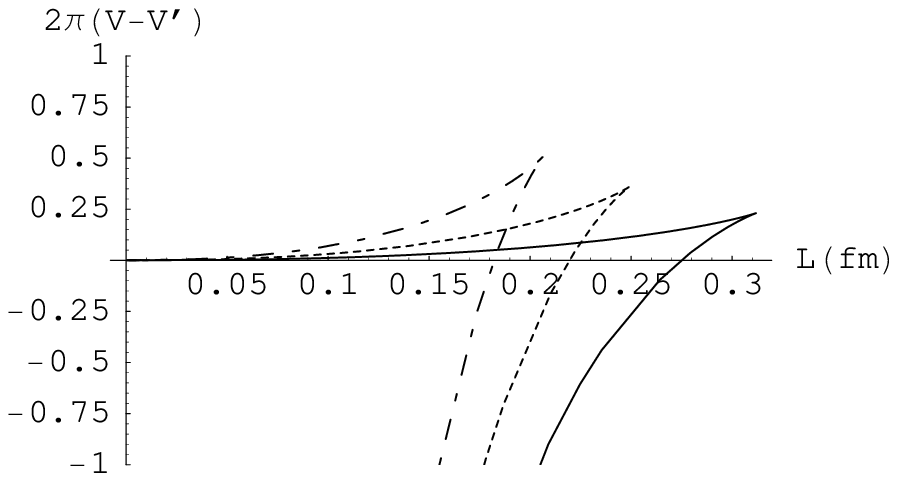}
\hfill
\includegraphics[width=3.2in]{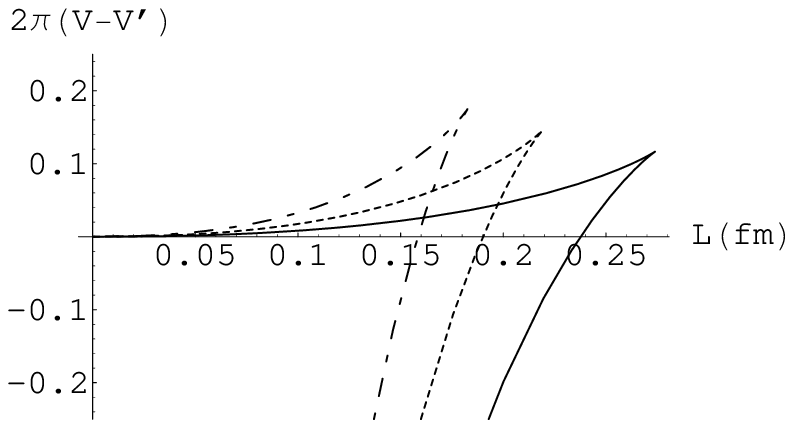}
\caption{The heavy quark potential as a function of the distance between quark and antiquark for the metric $n=3$~(left) and $n=4$~(right).  The solid line is for $T=0.20$ GeV.  The dashed line is for $T=0.25$ GeV.  And the dashed-dot line is for $T=0.30$ GeV.}
\label{add1-fig}
\end{figure}

The total potential between quark and antiquark is then given by
\begin{eqnarray}
V-V^{\prime}.
\end{eqnarray}
The plots between $V-V^{\prime}$ and $L$ for both metrics are represented in Fig.~{\ref{add1-fig}}.  The screening length of the quark-antiquark configuration is the distance $L$ when $V-V^{\prime}=0$ beyond which the parallel strings configuration has lower energy and more stable.  The plots between $V-V^{\prime}$ and $L$ for both metrics are shown in Fig.~\ref{add1-fig}.  In the phase of quark-gluon plasma at certain temperature, the mesonic state being submerged into the plasma will tend to ``melt'' when the screening length is smaller than the natural orbital radius of the meson. 

For $R_{n}=1, u_{max}=20, T=0.25~(0.20, 0.30)$ GeV, we find that the screening length, $L^{*}$ is 

\underline{for $n=3$},
\begin{eqnarray}
L^{*}=0.22~(0.275, 0.182)\, ~\mbox{fm}, \label{eq:L1}
\end{eqnarray}

\underline{for $n=4$},
\begin{eqnarray}
L^{*}=0.19~(0.24, 0.158)\, ~\mbox{fm}.  \label{eq:L2}
\end{eqnarray}
The relationship can be represented as  
\begin{eqnarray}
L^{*} & = & 0.055~(0.0475)\times \frac{\mbox{ fm GeV}}{T}  \quad\quad\quad\mbox{ for $n=3~(4)$}. \label{eq:LT}
\end{eqnarray}\\
Note that this $1/T$ dependence is exact in the $u_{max}\to \infty$ limit for {\it both metrics} and the coefficient for each $n$ can be calculated numerically.  Moreover, in this $u_{max}\to \infty$ limit, the screening lengths for both metrics are independent of $R$.  Introduction of the finite cutoff through position of the probe brane $u_{max}$ brings in soft $R$-dependence of the screening lengths.  However, we will set $R=1, u_{max}=20$ throughout our analysis hereafter.

\subsection{$\omega\neq0$ case}

We incorporate the angular momentum of the quark-antiquark system by rotating the metric according to the {\it ansatz} Eqn.~(\ref{eq:g1}).  For $\omega\neq0$, we have the potential containing extra term $-\omega^{2}\rho^{2}$,
\begin{eqnarray}
V=\frac{1}{\pi}\int d\rho\sqrt{\left(\frac{u^{\prime2}}{f_{n}\left(u\right)}+\left(\frac{u}{R_{n}}\right)^{n}\right)\left(f_{n}\left(u\right)-\omega^{2}\rho^{2}\right)},
\label{eq:V1}
\end{eqnarray}

the angular momentum then can be calculated from $\partial L/\partial \omega$,
\begin{eqnarray}
J=\int d\rho\frac{\omega\rho^{2}\left(\frac{u^{\prime2}}{f_{n}\left(u\right)}+\left(\frac{u}{R_{n}}\right)^{n}\right)}{\sqrt{\left(\frac{u^{\prime2}}{f_{n}\left(u\right)}+\left(\frac{u}{R_{n}}\right)^{n}\right)\left(f_{n}\left(u\right)-\omega^{2}\rho^{2}\right)}}.
\label{eq:J1}
\end{eqnarray}

Again with the following gauge for the parallel strings configuration
\begin{eqnarray}
t=\tau,\,\,\,u=\sigma,\,\,\, \rho=\rho_{0},\,\,\, \varphi=\omega t,
\label{eq:g2}
\end{eqnarray}
the corresponding regulating potential for the parallel strings configuration becomes
\begin{eqnarray}
V^{\prime}=\frac{1}{\pi}\int_{u_{c}}^{u_{max}}\sqrt{1-\frac{\rho_{0}^{2}\omega^{2}}{f_{n}(u)}} ~du
\end{eqnarray}
where $u_{c}=u_{h}/(1-\rho_{0}^{2}\omega^{2})^{1/n}$, the minimum distance in the $u$-direction for rotating parallel strings.  It is interesting to note that in the rotating metric, the position of horizon where parallel strings end is shifted from $u_{h}$ to $u_{c}=u_{h}/(1-\rho_{0}^{2}\omega^{2})^{1/n} > u_{h}$.  

We can numerically solve the equation of motion Eqn.~(\ref{eom}) for given value of $\omega,u_{max}$, and numerically estimate the potentials and angular momentum.  

\begin{figure}[angle=270]
\centering
%\epsfxsize=4.0in
%\hspace*{0in}
%\epsffile{screening302and025and03.ps}
\includegraphics[height=5in, angle=270]{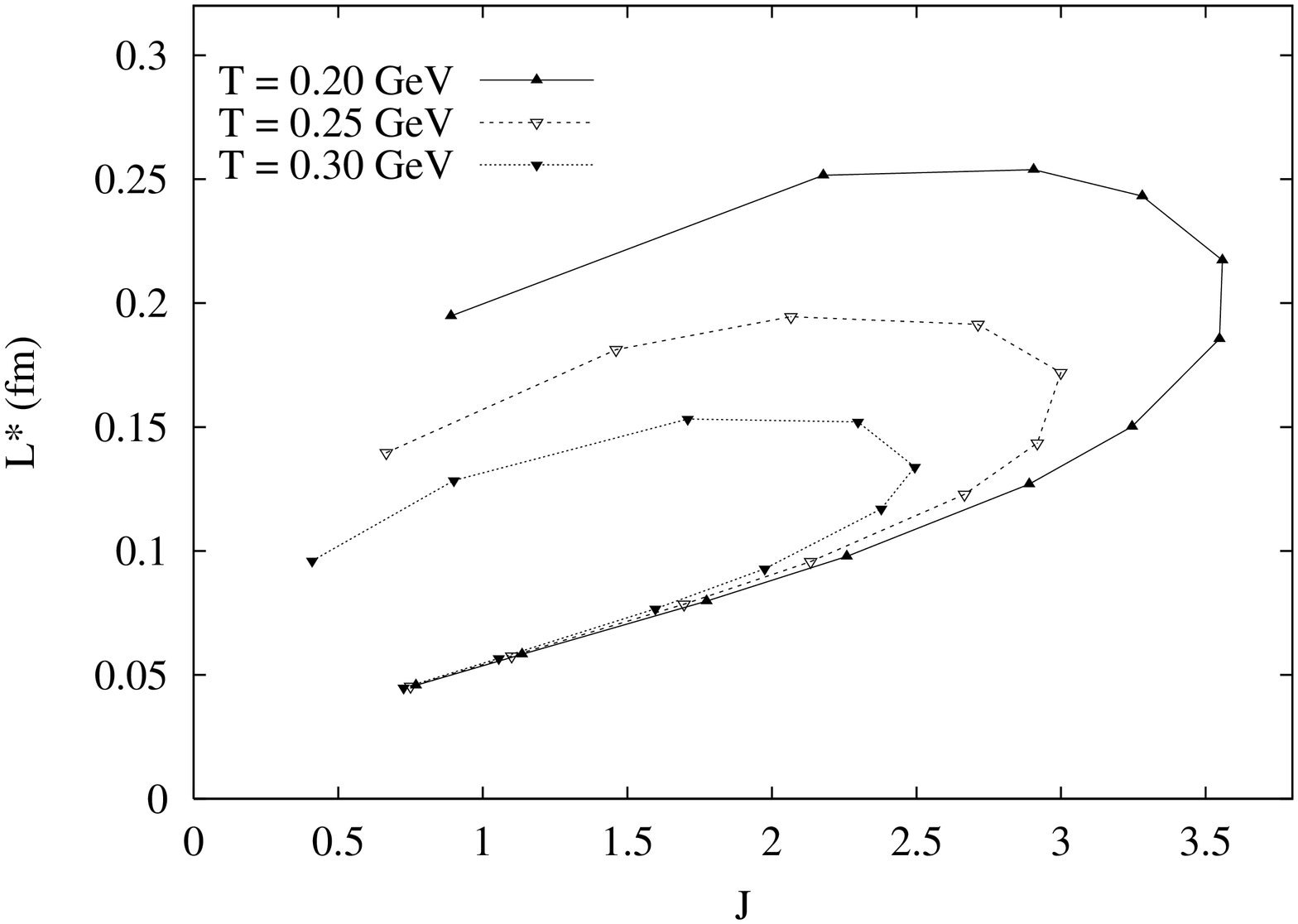}
\caption{The screening length of U-shaped strings in the metric $n=3$, as a function of the angular
momentum $J$ at $T= 0.20, ~0.25, ~0.30$ GeV.  From above in the clockwise direction, $\omega$ starts from the value around 0.2-0.3 and increases to around 8 for each temperature curve. }
\label{1-fig}
\end{figure}

For $n=3, T= 0.20, ~0.25, ~0.30$ GeV, we can numerically estimate the screening length
for each angular momentum where $V-V^{\prime}=0$ as in Figure \ref{1-fig}.  The upper part of the curves where values of $\omega$ are small corresponds to the physical region approximating mesonic states between quark and antiquark in the gauge plasma.  Motivated by Eqn.~(\ref{eq:LT}) for $\omega=0$ non-rotating case, we fit the relationship between the screening length $L^{*}$ and $T$ at fixed angular momentum $J$ numerically as
\begin{eqnarray}      
L^{*}(n=3) & \simeq & 0.0406~(0.0496)\times \frac{\mbox{ fm GeV}}{T}  \quad\quad\quad\mbox{ for $J=1~(2)$}. \label{eq:Lw3}
\end{eqnarray}

\begin{figure}[tbp]
\centering
%\epsfxsize=4.0in
%\hspace*{0in}
%\epsffile{screening402and025and03.ps}
\includegraphics[height=5in, angle=270]{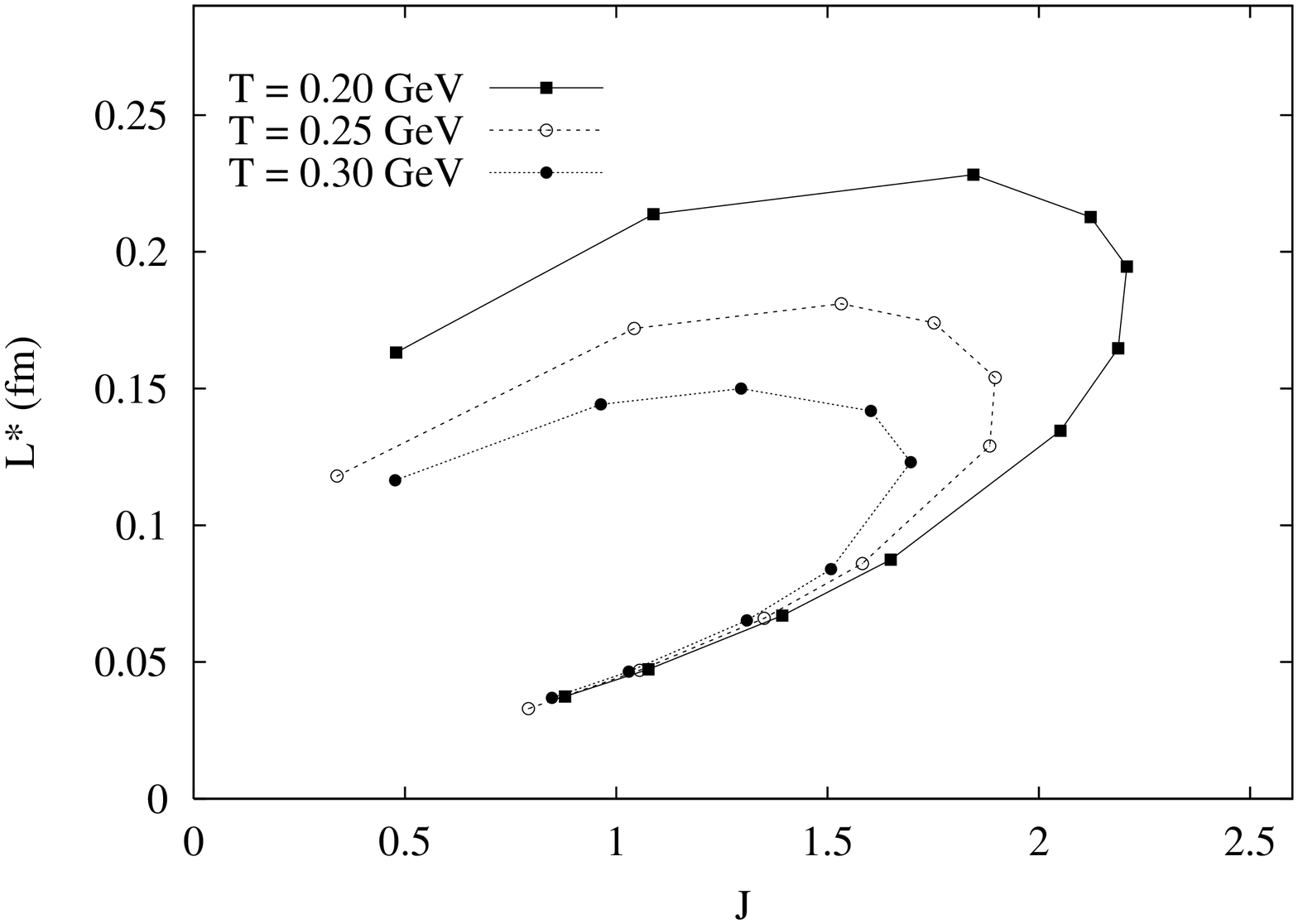}
\caption{The screening length of U-shaped strings in the metric $n=4$, as a function of the angular
momentum $J$ at $T=0.20, ~0.25, ~0.30$ GeV.  From above in the clockwise direction, $\omega$ starts from the value around 0.2-0.3 and increases to around 8 for each temperature curve. }
\label{2-fig}
\end{figure}

For $n=4, T= 0.20, ~0.25, ~0.30$ GeV, we can numerically estimate the screening length
for each angular momentum as in Figure \ref{2-fig}.  The $1/T$ fit for this case is 
\begin{eqnarray}      
L^{*}(n=4) & \simeq & 0.0424~(0.0452)\times \frac{\mbox{ fm GeV}}{T}  \quad\quad\quad\mbox{ for $J=1~(2)$}. \label{eq:Lw4}
\end{eqnarray}
The $1/T$ dependence of the screening length $L^{*}$ for each dual metric with and without rotation is summarized in Table~\ref{t1}.

\begin{table}[tb]
{\tabcolsep=.5cm  %% adjust space between columns
\def\arraystretch{1.5}  %% adjust space between rows
\medskip
\centering
\begin{tabular}{|c|c|c|c|} \hline
$n $&  $J=0~(\omega=0)$  &  $J=1$  &  $2$  \\ \hline
$3$ & $0.055$  &  $0.0406$ & $0.0496$  \\
$4$ & $0.0475$  &  $0.0424$ & $0.0452$  \\ \hline
\end{tabular}  }
\caption[]
{The values of $L^{*}T$ in unit of (fm$\times$ GeV) for the dual metric $n=3, 4$ with and without the rotation. } \label{t1}
\end{table}

\begin{figure}[tbp]
\centering
%\epsfxsize=4.0in
%\hspace*{0in}
%\epsffile{screening3025and4025.ps}
\includegraphics[height=5in, angle=270]{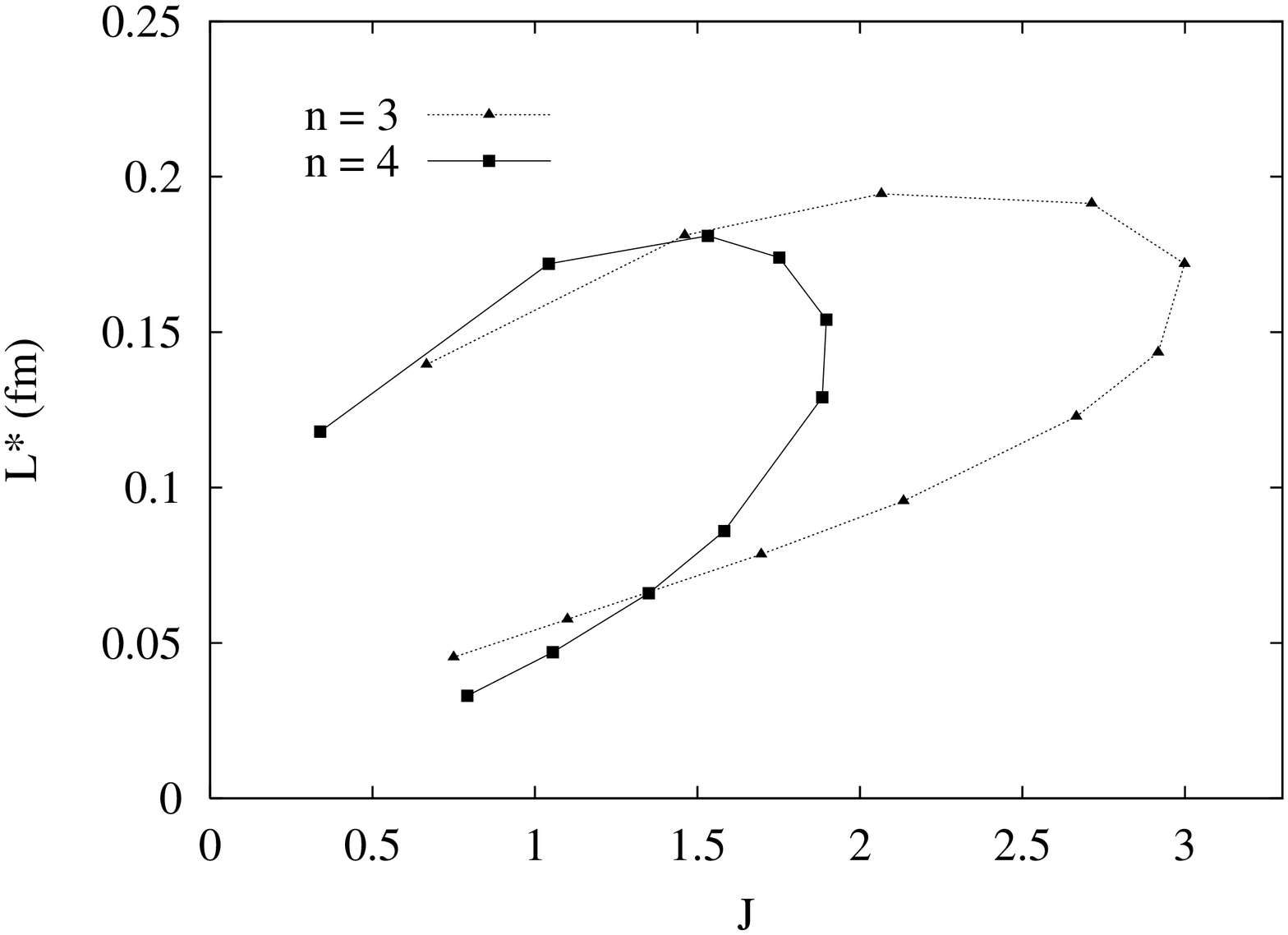}
\caption{The screening length for $n=3$ and $n=4$ at $T=0.25$ GeV. }
\label{3-fig}
\end{figure}

\begin{figure}[p]
\centering
%\epsfxsize=4.0in
%\hspace*{0in}
%\epsffile{crittemp30and40.ps}
\includegraphics[height=5in, angle=270]{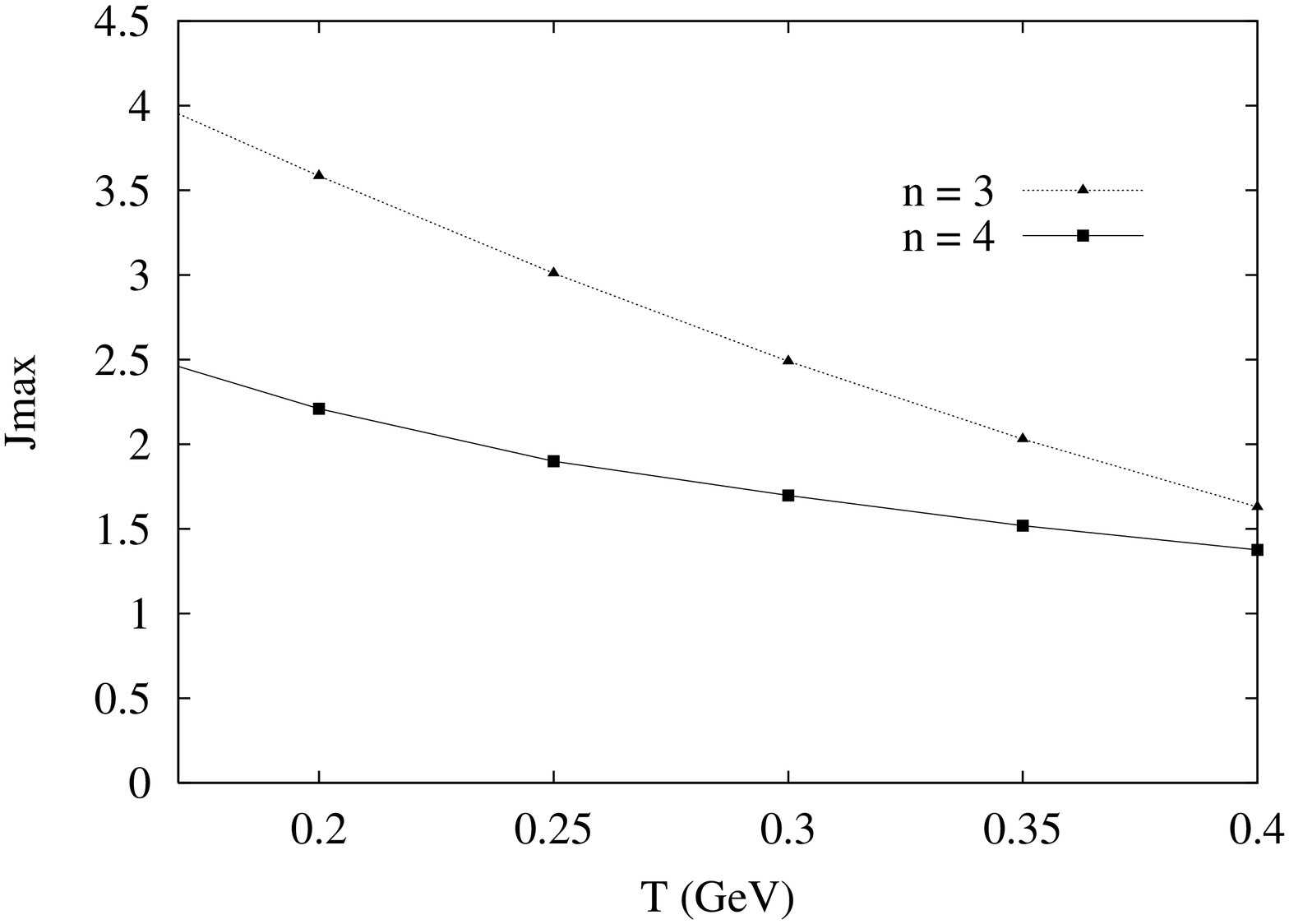}
\caption{The maximal spin $J$ as functions of the temperature for metric $n=3, 4$.  Mesons with higher spin will melt at each corresponding temperature. }
\label{4-fig}
\end{figure}

We can compare the screening length for $n=3$ and $n=4$ when
the temperature $T=0.25$ GeV as presented in Figure \ref{3-fig}.  The $n=4$ metric covers smaller range of values of $J$ at each temperature.  This could also be seen in Fig.~\ref{4-fig}.  Note that the relationships Eqn.~(\ref{eq:Lw3}),(\ref{eq:Lw4}) are valid only up until the temperatures where the quark antiquark system melts as is shown in Fig.~\ref{4-fig}.

%\begin{figure}[p]
%\centering
%\epsfxsize=4.0in
%\hspace*{0in}
%\epsffile{screening3and4.ps}
%\caption{The screening length for $n=3$ and $n=4$ at $T=0.20, 0.25, 0.30$ GeV. }
%\label{3.5-fig}
%\end{figure}

Mesons with particular angular momentum $J$ in the gauge plasma will melt at different temperature.  What we mean by ``melting'' is that regardless of the distance between quark and antiquark, it is always preferred energetically for the pair to be separated as free states.  In the dual string picture, this corresponds to the situation when the energy of the parallel strings is lower than the connected rotating configuration.  We define the maximal spin $J_{max}$ at each temperature to be the value of spin below which connected-string configuration has {\it finite} screening length, and above which the parallel strings configuration is always energetically preferred at any distance.  

From Fig.~\ref{1-fig} and Fig.~\ref{2-fig}, the values of $J_{max}$ are the ones at the tip of each temperature curve.  It should be emphasized that this is different from the $J_{max}$ derived from the relations between $E^{2}$~($E$ is total energy of the strings configuration) and $J$ as in Ref.~\cite{psz}.  The mesons actually melt before we reach the spike in the plot $E^{2}$ and $J$.  In other words, it is the regulated potential $V-V^{\prime}$~(i.e. difference between energy of connected-string and parallel strings configurations) which indicates the melting of mesons, not the total energy $E$. 
    
Since the charm~($m_c\simeq1.3$ GeV) and bottom~($m_b\simeq4.7$ GeV) quarks are heavy, the meson spectroscopy can be estimated based on the non-relativistic potential given by Eqn.~(\ref{eq:clpot1}).  From Ref.~\cite{kms}, the screening radius between quark and antiquark~($r_{D}\equiv 1/m_{D}$) in $J/\psi$ is around $0.283$ fm.  If we naively identify the angular momentum of the classical string $J$ with the orbital angular momentum $\ell$~(instead of $\sqrt{\ell(\ell +1)}$~) of the mesonic state, this value of the screening radius will correspond to the screening lengths $L^{*}$ calculated in the gravity duals from Eqn.~(\ref{eq:LT}) at $T=194,~168$ MeV for $n=3,~4$ respectively.  For $b \bar{b}$ mesonic states~($\Upsilon,\chi_{b}$, etc.), the screening radius between quark and anti-quark~($r_{D}$) in $\Upsilon$ is $0.127$ fm, corresponding to the screening lengths $L^{*}$ at $T=433,~374$ MeV for $n=3,~4$ metric.  We summarize, using Eqn.~(\ref{eq:LT}),(\ref{eq:Lw3}), and (\ref{eq:Lw4}), the corresponding temperature $T_{D}$ that our screening length gives the same value as the screening radius for each quarkonium from Ref.~\cite{kms} in Table~\ref{t2}.  

\begin{table}[tb]
{\tabcolsep=.5cm  %% adjust space between columns
\def\arraystretch{1.5}  %% adjust space between rows
\medskip
\centering
\begin{tabular}{|c|c|c|c|c|} \hline
&$\Upsilon~(\Upsilon^{\prime})~(\ell=0)$&$\chi_{b}~(\ell=1)$& $J/\psi~(\psi^{\prime})~(\ell=0)$&$\chi_{c}~(\ell=1)$ \\
 \hline
$T_{D}/ \mbox{MeV}~(n=3)$&$433~(187)$&$115$&$194~(99)$&$70$\\
$T_{D}/ \mbox{MeV}~(n=4)$&$374~(161)$&$119$&$168~(85)$&$73$ \\ \hline
\end{tabular}  }
\caption[]
{The corresponding melting temperature $T_{D}$ for different bound states where the screening length $L^{*}$ calculated in the gravity duals takes the value of the screening radius $r_{D}$ derived empirically in Ref.~\cite{kms}.  We identify the angular momentum of the classical string $J$ with the orbital angular momentum $\ell$ of the mesonic state. } \label{t2}
\end{table}
  
It must be emphasized that these numbers $T_{D}$ are naively estimated using screening length in the gravity duals to match with the screening radius from empirical fitting.  New results from lattice calculations~\cite{lat} show surprising resistance to melt of the 1S charmonium states such as $J/\psi, \eta_{c}$.  It does not seem plausible to us that this result can be understood in terms of the screening effect alone.   

\section{Drag Forces}

A few comments can be made regarding drag forces in both metrics.  We will only consider the case when $\omega=0$ and the metric is boosted along $x^{1}$ direction with the {\it ansatz} $u=\sigma$,
\begin{eqnarray} 
x^{1}(t,u) & = & vt + \rho(u).
\label{eq:d1}
\end{eqnarray}
The Lagrangian becomes
\begin{eqnarray}
L & = & \sqrt{1-\frac{v^{2}}{f_{n}(u)}+\frac{f_{n}(u)}{H(u)}\rho^{\prime2}}
\label{eq:d2}
\end{eqnarray}
where $H(u)\equiv R_{n}^{n}/u^{n}$.  As in Ref.~\cite{gub2} where the $n=4$ metric has been calculated, since the Lagrangian does not depend on $\rho$ explicitly, we can define the constant $\pi_{\rho}\equiv \partial L/\partial \rho^{\prime}$.  The equation of motion then becomes
\begin{eqnarray}
\rho^{\prime} & = & \pi_{\rho}\frac{H}{f}\sqrt{\frac{f_{n}-v^{2}}{f_{n}-\pi_{\rho}^{2}H}}, \\
\pi_{\rho}    & = & \frac{v}{\sqrt{1-v^{2}}}\left( \frac{u_{h}}{R_{n}} \right)^{n/2}
\end{eqnarray}
where the latter is chosen to guarantee the reality of $\rho^{\prime}$.  The drag force is given by
\begin{equation}
\frac{dp_{1}}{dt} = -\frac{\sqrt{1-v^{2}}}{2\pi \alpha^{\prime}}G_{x^{1}\nu}g^{u\alpha}\partial_{\alpha} X^{\nu} = -\frac{1}{2\pi\alpha^{\prime}}\frac{v}{\sqrt{1-v^{2}}}\left( \frac{u_{h}}{R_{n}} \right)^{n/2} \\
\end{equation}
\begin{equation}
\frac{dp_{0}}{dt} = -\frac{\sqrt{1-v^{2}}}{2\pi \alpha^{\prime}}G_{x^{0}\nu}g^{u\alpha}\partial_{\alpha} X^{\nu} = -\frac{1}{2\pi\alpha^{\prime}}\frac{v^{2}}{\sqrt{1-v^{2}}}\left( \frac{u_{h}}{R_{n}} \right)^{n/2}. \\
\end{equation}
Note that the energy loss
\begin{eqnarray}
\frac{dp_{0}}{dt} = v\frac{dp_{1}}{dt}.
\end{eqnarray}
For $n=3, 4$; the string tension is given by
\begin{eqnarray}
\frac{1}{2\pi \alpha^{\prime}}& = & \frac{\lambda_{5}}{27\pi^{2}R_{3}^{3}},\quad \frac{\sqrt{\lambda_{4}}}{2\pi R_{4}^2};
\end{eqnarray}
the drag force is
\begin{equation}
\frac{dp_{1}}{dt} = -\frac{v}{\sqrt{1-v^{2}}}\left\{
\begin{array}{l}
(\frac{4}{9})^{3}\pi(\lambda_{5}T^{3}) \qquad ~~\mbox{  for $n=3$},  \\ [3mm]
\frac{\pi}{2}(\sqrt{\lambda_{4}}T^{2}) \qquad \quad~\mbox{  for $n=4$}.
\end{array}
\right.
\end{equation}
The 't~Hooft coupling $\lambda_{5}\equiv g_{5}^{2}N=(2\pi)^{2}g_{s}l_{s}N$ and $\lambda_{4}\equiv g_{YM}^{2}N=4\pi g_{s}N$ respectively.  The corresponding relaxation times are
\begin{displaymath}
t_{0} = \left\{
\begin{array}{l}
(\frac{9}{4})^{3}(\frac{1}{\pi})(m/\lambda_{5}T^{3}) \qquad ~~~~\mbox{  for $n=3$},  \\ [3mm]
(\frac{2}{\pi})(m/\sqrt{\lambda_{4}}T^{2}) \qquad \qquad ~\mbox{  for $n=4$}.
\end{array}
\right.
\end{displaymath}

Then the relaxation time for the charm quark 
\begin{eqnarray}
t_{0}& = & 0.22~\mbox{fm/c}~\frac{m/m_{c}}{(g_{5}^{2}N/50\pi~\mbox{GeV}^{-1})(T/300~\mbox{MeV})^{3}} \qquad \qquad ~\mbox{  for $n=3$},  \\ [3mm]
t_{0}& = & 0.58~\mbox{fm/c}~\frac{m/m_{c}}{\sqrt{g_{YM}^{2}N/10}(T/300~\mbox{MeV})^{2}} \qquad \qquad \qquad ~~~\mbox{  for $n=4$}.
\end{eqnarray}
The relaxation time for the bottom quark
\begin{eqnarray}
t_{0}& = & 0.80~\mbox{fm/c}~\frac{m/m_{b}}{(g_{5}^{2}N/50\pi~\mbox{GeV}^{-1})(T/300~\mbox{MeV})^{3}} \qquad \qquad ~\mbox{  for $n=3$},  \\ [3mm] 
t_{0}& = & 2.08~\mbox{fm/c}~\frac{m/m_{b}}{\sqrt{g_{YM}^{2}N/10}(T/300~\mbox{MeV})^{2}} \qquad \qquad \qquad ~~~\mbox{  for $n=4$}.
\end{eqnarray}

The drag force and relaxation time for the metric $n=3$ are more sensitive to the temperature than the metric $n=4$ in contrast to the screening length which has the same temperature dependence $1/T$ for both metrics.  The reason is the dependence on the string tension $1/2\pi \alpha^{\prime}$ and consequently on the 't~Hooft coupling $\lambda_{p+1}~(p=7-n)$ of the drag force which has nonzero mass dimension for the gauge theory in spacetime with dimension higher than 4~(i.e. $[\lambda_{5}]= length$).  For the screening length, the weak $R$-dependence~(and therefore $\lambda$-dependence) is only induced by introducing finite position of the probe branes $u_{max}$.  In the $u_{max}\to \infty$ limit, the screening length becomes independent of $R$~(and $\lambda$).  As a consequence, we have the unique $1/T$ dependence~(with however, different coefficients) of the screening length in both dual metrics $n=3$ and $4$ in contrast to the drag force.

\begin{figure}[p]
\centering
\includegraphics[height=5in, angle=270]{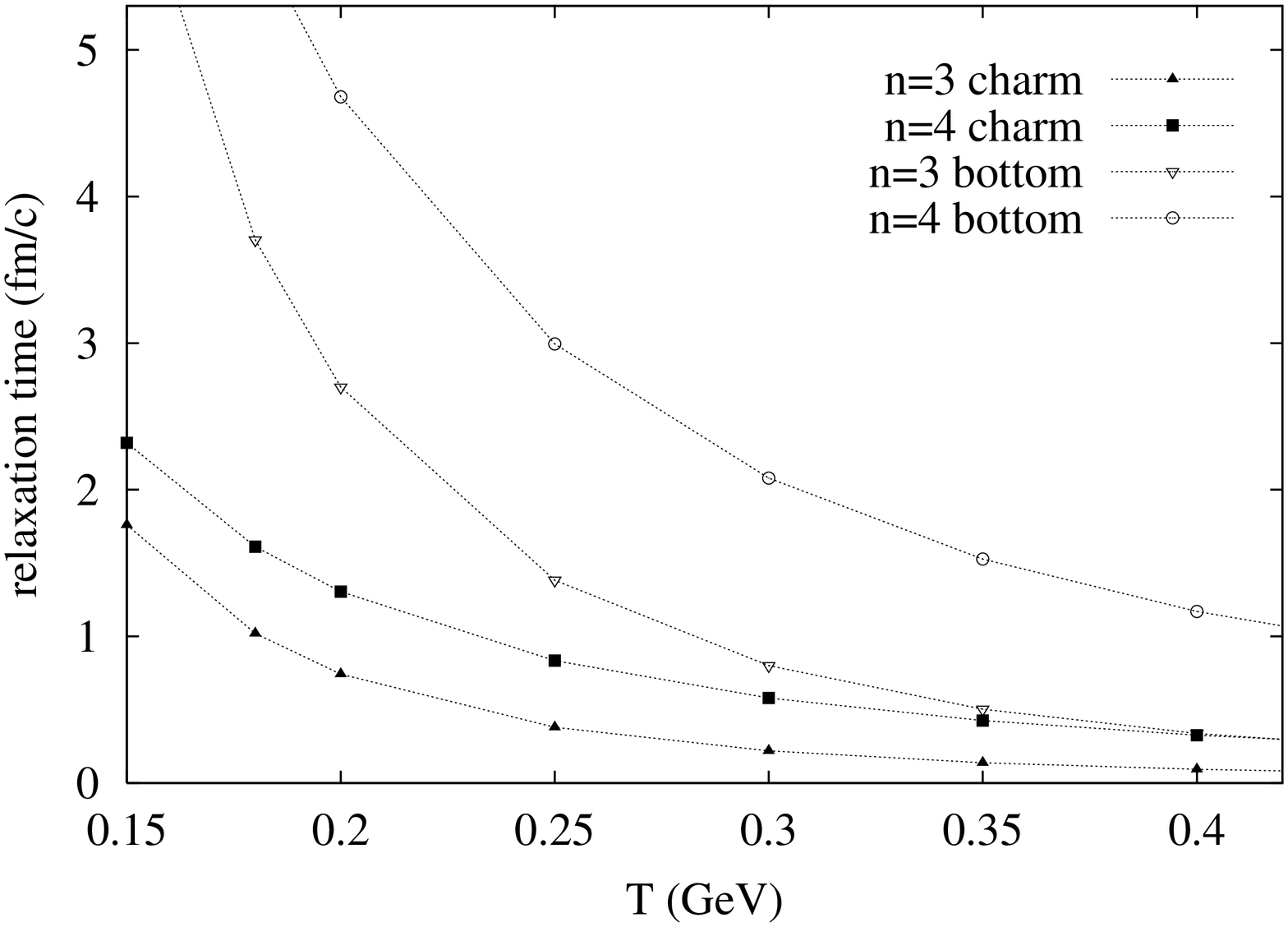}
\caption{The relaxation time versus temperature for charm and bottom from the metric $n=3,~4$.  Note the similar apparent trend of decreasing of the relaxation time with the temperature between the two metrics for each quark.  The values of $\lambda_{4},\lambda_{5}$ are set to be $10,50\pi$ GeV$^{-1}$ respectively. }
\label{5-fig}
\end{figure}

It is interesting to note that even though $n=3$ metric gives shorter relaxation times than the metric $n=4$, and despite the different analytic $T$-dependence, the plotted temperature dependence of $t_{0}$ between the two metrics as shown in Fig.~\ref{5-fig} appear very similar to each other for each case of the quark.  Comparing these values at RHIC and LHC~(higher temperature) will reveal which metric is more suitable as the gravity dual of the QCD, in addition to the screening length, Eqn.~(\ref{eq:L1}) and (\ref{eq:L2}), which seems to favor the metric $n=3~(L^{*}=0.22$ fm) when compared to the lattice result $r_{D}\simeq 0.21\sim 0.24$ fm at $T=250$ MeV for $T_{c}=175\sim 190$ MeV~\cite{kz}.  

Calculation of the jet quenching parameters in Sakai-Sugimoto metric~\cite{gxz} has given the similar results confirming the non-universal behaviour of the drag forces and the jet quenching with more sensitivity to the temperature.  A similar calculation of the drag force in the $n=3$ metric has also been done in Ref.~\cite{tal} without mentioning the relaxation time and its temperature dependence.  We emphasize that this non-universal behaviour of the drag forces and energy losses is due to the dependence on the 't~Hooft coupling of the quantities we are considering.  

\section{Conclusions and Discussions}

We have considered two gravity duals of the non-Abelian gauge theory in 1+3 dimension~(after compactified the extra dimension).  One is Sakai-Sugimoto model and another is the usual near-horizon metric $AdS_{5}$.  Both metrics have horizons which induce scales and temperature into the dual theories.  

The angular momentum is incorporated into the analysis by rotation of the background metrics.  This provides us with a quantitative way to calculate the screening lengths for varying angular momentum, as well as the temperature dependences of the screening length for non-zero angular momentum states.  An interesting result is the fact that $1/T$ dependence of the screening lengths appears to be valid with good accuracy even for non-zero angular momentum states.  

A naive comparison~(Table~\ref{t2}) to the melting of physical quarkonium states is performed and shown that both gravity duals, at least in the rotated approach, cannot explain the surprising lattice results on the resistance to melt of the 1S charmonium states such as $J/\psi$ and $\eta_{c}$~\cite{lat}.  This suggests that the realistic melting of quarkonium is more complicated and could not probably be explained by consideration of the plasma after its formation alone in terms of the screening radius.  However, the non-1S states such as $\psi^{\prime},\chi_{c}$ seem to melt away both in lattice and by consideration of the screening length.  It is not unnatural to wonder the same phenomenon would repeat for the bottomonia where $\Upsilon^{\prime},\chi_{b}$ melt but $\Upsilon$ resist when submerged into the quark-gluon plasma.  

The drag forces in both gravity duals share the same aspect that they depend on positive power of the 't~Hooft couplings~($\lambda_{5},\sqrt{\lambda_{4}}$).  Theoretically this shows that the relaxation times vanish in the large 't~Hooft limit $\lambda\to \infty$, in other words, the quark-gluon plasma becomes extremely resistive against the quarks passing through.  This is in contrast to the shear viscosity per entropy density value which approaches finite constant in the large $\lambda$ limit due to the fact that it does not depend on the 't~Hooft coupling at the leading order~(for $AdS_{5}$).  Interestingly, the screening length in the $u_{max}\to \infty$ limit for $J=0$ also share the same characteristic that it does not depend on the 't~Hooft coupling, and thus shows the same temperature dependence $1/T$ in both gravity duals.

Even though some phenomena such as the realistic ``melting'' of quarkonium are more complicated and require detailed understanding of the phase transitions in the QCD either from the lattice approach, or from other QCD methods, some ``universal'' aspects of the hot gauge plasma and the behaviour of quarkonium in the medium could be understood and even quantitatively predicted by analysis on the gravity duals of the strongly-coupled gauge theories.  We hope that the results in this article would reveal more or less certain aspects of the quark-gluon plasma as well as providing some hints into the exact nature of this nuclear phase to be produced at the upcoming LHC.

\section*{Acknowledgments}
\indent
P.B. was supported in part by the U.S. Department of Energy under contract number DE-FG02-01ER41155.  J.L was supported in part by the NSF Grant No. PHY-0340729.

\end{document}